\documentclass{elsart}
\usepackage{graphics}
\begin{document}
\begin{frontmatter}
\hyphenation{Coul-omb}
\title{Why Use a Hamilton Approach in QCD?}
\author{H. Kr\"oger, X.Q. Luo and K.J.M. Moriarty}
\address{
D\'epartement de Physique, Universit\'e Laval, Qu\'ebec, Qu\'ebec G1K 7P4, 
Canada \\ 
Department of Physics, Zhongshan University, Guangzhou 510275, China \\ 
Department of Mathematics, Statistics and Computer Science, 
Dalhousie University, Halifax, Nova Scotia B3H 3J5, Canada} 
\date{1 August 2000}
\begin{abstract} 
We discuss $QCD$ in the Hamiltonian frame work. We treat finite density $QCD$ in the strong coupling regime. We present a parton-model inspired regularisation scheme to treat the spectrum ($\theta$-angles) and distribution functions in $QED_{1+1}$. We suggest a Monte Carlo method to construct low-dimensionasl effective Hamiltonians. Finally, we discuss improvement in Hamiltonian $QCD$.
\end{abstract}
\end{frontmatter}

\noindent {\bf 1. Introduction} \\
There are presently the following problems in $QCD$: 
How to compute non-perturbatively (on the lattice):  
(a) $QCD$ at non-zero chemical potential - finite quark density. 
(b) Hadron structure functions, at small $x_{B}$ and small $Q$. 
(c) $S$-matrix, scattering cross sections, phase shifts,
(d) Decay amplitudes. 
Here, we want to discuss recent progress in the Hamiltonian formulation, 
$QCD$ at non-zero chemical potential in the strong coupling regime, 
distribution function in $QED_{1+1}$, 
the Monte Carlo Hamiltonian, 
and improvement of the continuum behavior of the Hamiltonian.

\bigskip

\noindent {\bf 2. Finite density QCD} \\
According to the big bang model of cosmology
creation of quarks and gluons occured at about $10^{-35}$ sec, 
quark-gluon plasma ($QGP$) was formed at about $10^{-6}$ sec, 
then there was a phase transition
and the formation of protons and neutrons began at $10^{-4}$ sec.
Today the universe is in a low temperature and low density phase: 
quarks are confined. 
Quark-gluon plasma may exist in cores of very dense stars 
like neutron stars. 
Experimental search for $QGP$ is carried out at the 
Relativistic Heavy Ion Collider (RHIC) at BNL and at the  
Large Hadron Collider (LHC) at CERN.
Theorists search for a precise determination of phase structure of $QCD$ 
at finite temperature $T$ and finite chemical potential 
$\mu$ (finite density). 
Shuryak \cite{Shuryak:00}
has predicted that $QCD$ at high temperature and density 
displays beyond the $QGP$ pase transition a much richer phase structure, 
due to strong instanton effects.

Here we suggest to consider finite temperature and finite density 
$QCD$ in the Hamiltonian formulation.
Consider the grand canonical partition function at temperature $T$ and chemical 
potential $\mu$ 
\begin{eqnarray}
Z &=& Tr e^{-\beta(H - \mu N)}, ~~~ 
N = \int d^{3}x ~ \psi^{\dagger}(x) \psi(x), ~~~
\beta = (k_{B} T)^{-1} .
\end{eqnarray}
$H$ is the Hamiltonian and $N$ is the fermion (quark) particle number operator,
No complex action problem occurs!
Let us consider as example
the energy density for free quarks 
\begin{equation}
\epsilon = \frac{1}{V} \frac{1}{Z} Tr H e^{-\beta(H-\mu N)} 
= - \frac{1}{V} \frac{\partial ln Z}{\partial \beta} |_{\mu, \beta} .
\end{equation}
Taking the zero-temperature limit ($T \to 0$) 
and the chiral limit ($m \to 0$) the energy density 
density (with the contribution $\mu=0$ subtracted) 
becomes
$\epsilon_{sub} = \frac{\mu^{2}}{4 \pi^{2}}$.
The same result is obtained by starting from a lattice Hamiltonian
with Wilson fermions.

\bigskip

\noindent {\bf Strong coupling QCD at non-zero chemical potential.} \\
Latttice $QCD$ at $\mu=0$ confines quarks and spontaneously breaks 
chiral symmetry.
Note: The strong coupling regime $1/g^{2} << 1$ is {\it not} compatible with 
continuum limit $a \to 0$.
The goal of the following calculation is to get a better 
understanding of mechanism of chiral phase transition. 
We make an ansatz for the lattice Hamiltonian in the strong coupling regime
following Ref.\cite{Luo:92}.
Then the following results are obtained \cite{Gregory:00}.
The vacuum energy becomes
\begin{eqnarray}
E_{\Omega} &=& 2 N_{f} N_{c} N_{s} 
\left[ (m^{(0)}_{dyn} - \mu) \Theta(\mu - m^{(0)}_{dyn})
- m^{(0)}_{dyn} -\mu \right] 
\nonumber \\
m^{(0)}_{dyn} &=& \frac{d}{a g^{2} C_{N}}, ~~~
C_{N} = (N_{c}^{2} -1)/(2N_{c}) ,  
\end{eqnarray}
where $m^{(0)}_{dyn}$ is the dynamical quark mass at 
$\mu=0$.
In order to compute the chiral condensate and the 
critical chemical potential
one can use the Feynman-Hellmann theorem.
The chiral condensate at $\mu = 0$ becomes
\begin{equation}
< \bar{\psi} \psi >^{(0)} = - 2 N_{C} [ 1 - 4 d/(g^{2} C_{N}^{2}) ] , 
\end{equation}
and the chiral condensate at $\mu \neq 0$ becomes
\begin{equation}
< \bar{\psi} \psi > = < \bar{\psi} \psi >^{(0)} 
\left[ 1 - \Theta(\mu - m^{(0)}_{dyn} ) \right] .
\end{equation}
As a result one has
\begin{eqnarray}
&& < \bar{\psi} \psi > \neq 0 ~ \mbox{for} ~ \mu < m^{(0)}_{dyn}: 
\mbox{chiral symmetry broken} ,
\nonumber \\ 
&& < \bar{\psi} \psi > = 0 ~ \mbox{for} ~ \mu > m^{(0)}_{dyn}:
\mbox{chiral symmetry restored} .
\end{eqnarray}
Thus we make the prediction of a first order chiral phase transition.
The critical chemical potential is
$\mu_{crit} = m^{(0)}_{dyn}$.
Our result for $\mu_{crit}$ is in agreement with the result obtained by 
Le Yaouanac et al.\cite{LeYaouanac:88}
However, Le Yaouanac et al. find a phase transition of 2nd order.
There have been predictions of the nucleon mass \cite{LeYaouanac:88}
$M_{nucl}^{(0)} \approx 3 m^{(0)}_{dyn}$.
Thus our result is
$\mu_{crit} \approx M_{nucl}^{(0)}/3$,
being consistent with other theoretical predictions \cite{LeYaouanac:88}.
We have moreover computed the quark number density, the corresponding susceptibility and some thermal masses \cite{Gregory:00}.

\bigskip

\noindent {\bf 3. Hamiltonian in a fast moving frame} \\
At hand of the massive Schwinger model, $QED_{1+1}$,  
we want to discuss the following topics:
(i) A physically motivated many-body method to solve a Hamiltonian
many-body problem and compute structure functions.
(ii) The treatment of the $\theta$-angle in Hamiltonian gauge theory.
The motivation for the Hamiltonian many-body method comes from 
the parton model
of deep inelastic lepton-hadron scattering. 
Experimentally, the Breit frame is a convenient choice for the 
interpretation of 
structure functions. 
We suggest to use a fast moving frame also in a computational study \cite{Kroger:98}. 
The Hamiltonian in axial gauge $A^{3}=0$ is given by
\begin{equation}
H = \int_{-L}^{L} dx^{3} ~ ( \bar{\psi} \gamma^{3} i \partial_{3} \psi + m \bar{\psi} \psi ) 
+ \frac{g^{2}}{2} \int_{-L}^{L} dx^{3} ~ (\psi^{\dagger} \psi) ~ \frac{1}{-\partial^{2}_{3}} 
~ (\psi^{\dagger} \psi ) .
\end{equation}
We start from a space-time lattice of spacing $a$ and length $[-L,L]$. 
Then we go over to a momentum lattice with cut-off $\Lambda=\pi/a$ and 
resolution $\Delta p = \pi/L$. 
We use the assumption (parton model): Left moving particles are not dynamically important,
when {\it physical} particles are considered from a reference frame with velocity 
$v = P/E \approx c$.
On the momentum lattice the following bounds hold:
$0 \leq p^{0}, p^{3} \leq P$.
A fast moving particle is Lorentz contracted. It fits into a small lattice volume
(compared to rest frame). 
A small lattice size implies 
a small number of virtual particle pairs created from vacuum: 
Number of vacuum pairs $\propto$ vacuum density $\times$ lattice size.
For the scaling window holds: 
$a \leq \mbox{Compton wavelength} \leq  L$ and 
$a \leq \xi a \leq L$,
where 
$\xi$ is the correlation length in dimensionless units
and $M = \frac{1}{\xi a}$ is the physical mass of particle in the ground state.
In the strong relativistic regime holds 
$M << P < \Lambda$.
In the limit $a \to 0$ the scaling window is replaced by:
$1/P \leq 1/M \leq L$. 
In Ref.\cite{Kroger:98} we have computed the mass spectrum: Vector boson and scalar boson, and compared it with predictions of chiral perturbation 
theory.
We also determined the dependence on the $\theta$ angle, defined by
a global gauge transformation 
$\Psi[A] \to e^{i n \theta} ~ \Psi[A], ~~ n=0, \pm 1, \pm 2, \dots$.
The $\theta$-action $S$ and topological charge $q$ are related by
$S[A,\theta] = S[A] - \theta q[A]$.  
In Ref. \cite{Kroger:98} we have also studied the massive Schwinger model and its dependence on the $\theta$-angle in the chiral limit. For the massless Schwinger model
the chiral condensate is analytically known,
\begin{equation}
\frac{ < \bar{\psi} \psi > }{ g } = \frac{ e^{\gamma} }{2 \pi \sqrt{\pi} } ~ \cos(\theta) 
\approx 0.1599 ~ \cos(\theta) .
\end{equation}
Using the Feynman-Hellmann theorem to relate the vector mass and the chiral condensate,
we determine the slope from numerical data and obtain
$\frac{ < \bar{\psi} \psi > }{ g } \approx 0.16 ~ \cos(\theta)$,
which is close to the analytical result.

The Hamiltonian approach also allows to compute wave functions, 
giving information on 
the parton structure.
\begin{equation}
f(x_{B}) = < \Psi_{V}(P) | \frac{1}{2} [ b^{\dagger}_{k} b_{k} 
+ d^{\dagger}_{k} d_{k} | \Psi_{V}(P) >,
\end{equation}
where
$x_{B} = \frac{k}{P}$ is the fraction of momentum of vector boson
carried by parton.
One observes that a large lattice (N=200) is needed
to resolve distribution functions precisely (see Ref.\cite{Kroger:98}).

\bigskip

\noindent {\bf 4. Stochastic Hamiltonian} \\
Lagrangian lattice field theory and its non-perturbative numerical solution 
by Monte Carlo methods is very successful. 
The typical number of configurations (depending on the statistical error) 
is in the order of $\sim 10^{2}$.
In Ref.\cite{Jirari:99,Huang:99}, we have propopsed to
construct an effective Hamiltonian, by 
treating the many-body problem by applying as far as possible 
Monte Carlo with importance sampling (e.g. Metropolis algorithm)
in analogy to Lagrangian lattice simulations. 
Let us consider as example 1-D Q.M. 
We compute transition matrix elements in imaginary time
\begin{equation}
M_{ij}(T) = <x_{i}| e^{-H T/\hbar} | x_{j} >, ~~~ i,j \in 1,\dots,N, 
\end{equation}
where $x_i$ are discrete position eigen states. We define 
an effective Hamiltonian via eigenfunctions and eigenvalues:
\begin{equation}
M(T) = U^{\dagger} D(T) U, ~~   
U^{\dagger}_{ik} = < e_{i}|E^{eff}_{k} >, ~~
D_{i}(T) = e^{-E^{eff}_{i} T/\hbar } .
\end{equation}
We compute matrix elements using 
Monte Carlo with importance sampling (e.g. Metropolis algorithm). 
Splitting the action
$S = S_{0} + S_{V}$, 
we treat $O=exp[-S_{V}/\hbar]$ as observable and compute the matrix element by 
Monte Carlo
\begin{eqnarray}
M_{ij}(T) = M_{ij}^{(0)}(T) ~
&\times& \int [dx] ~ exp[ -S_{V}[x]/\hbar] ~ 
exp[-S_{0}[x]/\hbar ] |_{x_{j},0}^{x_{i},T} 
\nonumber \\
&/& \int [dx] ~ exp[ -S_{0}[x]/\hbar] |_{x_{j},0}^{x_{i},T} .
\end{eqnarray}
The construction of a stochastic basis has been proposed in Ref.\cite{Huang:99}. The result of a numerical study is shown in Fig.[1].
\begin{figure}
\caption{Harmonic oscillator. Specific heat $C(\beta)$ from Monte Carlo Hamiltonian, using stochastic basis.}
\label{fig:Fig1}
\end{figure}

\bigskip

\noindent {\bf 5. Improvement in Hamiltonian QCD} \\
Szymanzik \cite{Szymanzik:83} suggested improvement 
program for lattice
field theory: Construction of "improved" action, 
such that error due to finite 
lattice spacing $a$ becomes smaller. 
Progress in improvement for lattice pure gauge theory was introduced 
by Lepage et al. \cite{Lepage:93},
suggesting tadpole improvement.
We have suggested how to construct an improved Hamiltonian 
on the lattice.
For pure gauge theory the 
construction of an improved Hamiltonian for $SU(3)$ gauge theory
via the transfer matrix has been reported by  
by Luo et al. in Ref.\cite{Luo:99}.
Improvement of the fermionic sector of a lattice Hamiltonian
has been carried out for 
$QCD_{1+1}$ by 
Luo et al. in Ref.\cite{Luo:94}. 
A numerical study was carried out by Jiang et al. \cite{Jiang:99}.
An example is shown in Fig.[2,3], showing that improvement reduces the dependence on Wilson's r-parameter.
\begin{figure}
\caption{$QCD_{1+1}$. Fermionic improvement.
$\chi=-\langle \bar{\psi} \psi \rangle_{sub}/(g_{lat}N_c)$ 
versus $1/g_{lat}^{2}$ for $N_C=3$ with Wilson fermions. 
Crosses: $r=0.1$, Diamonds: $r=1$.}
\label{fig:Fig2}
\end{figure}
\begin{figure}
\caption{$QCD_{1+1}$. Fermionic improvement.
$\chi=-\langle \bar{\psi} \psi \rangle_{sub}/(g_{lat}N_c)$ versus $1/g_{lat}^{2}$ for $N_C=3$ with improved Wilson fermions. 
Crosses: $r=0.1$, Diamonds: $r=1$.}
\label{fig:Fig3}
\end{figure}

\bigskip

\noindent {\bf 6. Conclusion} \\
Hamiltonian formulation of QCD is attractive, for the following reasons: 
(a) There is no sign or complex action problem! 
(b) It is suitable for the computation of wave functions and structure 
functions. 
(c) It is suitable for the computation of the S-matrix. 
The problem is: How to solve the Hamiltonian many-body system? 
(d) Proposal: Construct an effective Hamiltonian via Monte Carlo. 
(e) Improvement program: Also possible in Hamiltonian formulation.

\noindent {\bf Acknowledgement}.
H. Kr\"oger and K.J.M. Moriarty are grateful for support by NSERC(Canada).
X.Q. Luo has been supported by NNSF(China) and NSF(China) for Distinguished Young Scholars.

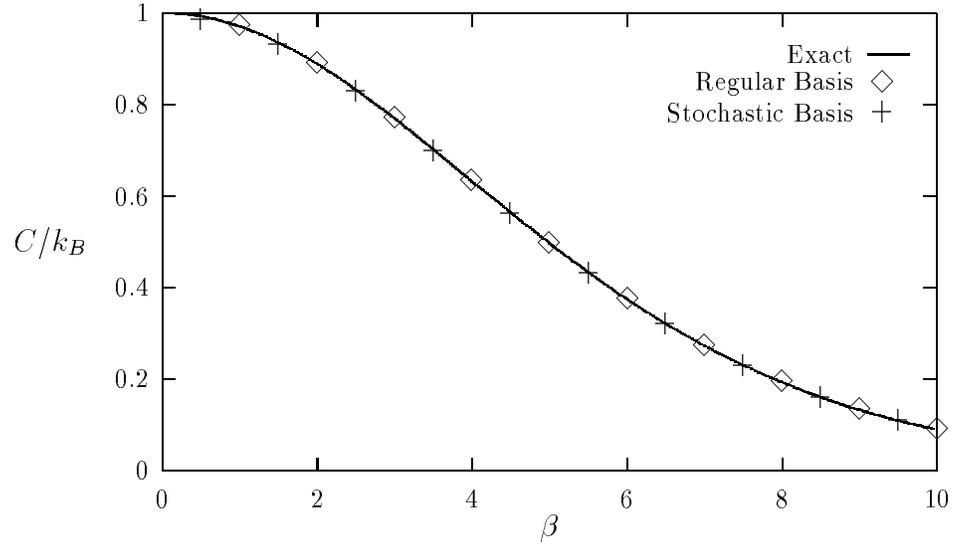
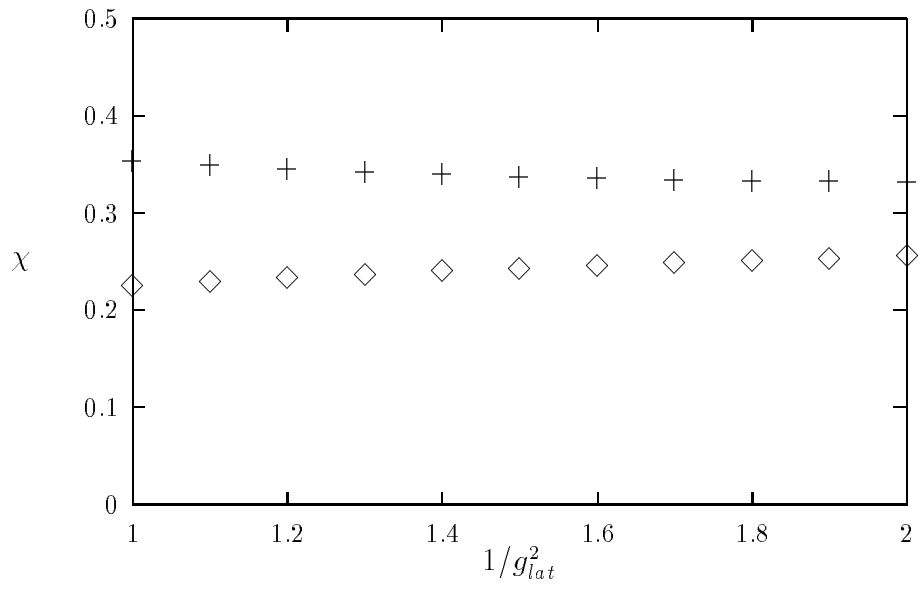
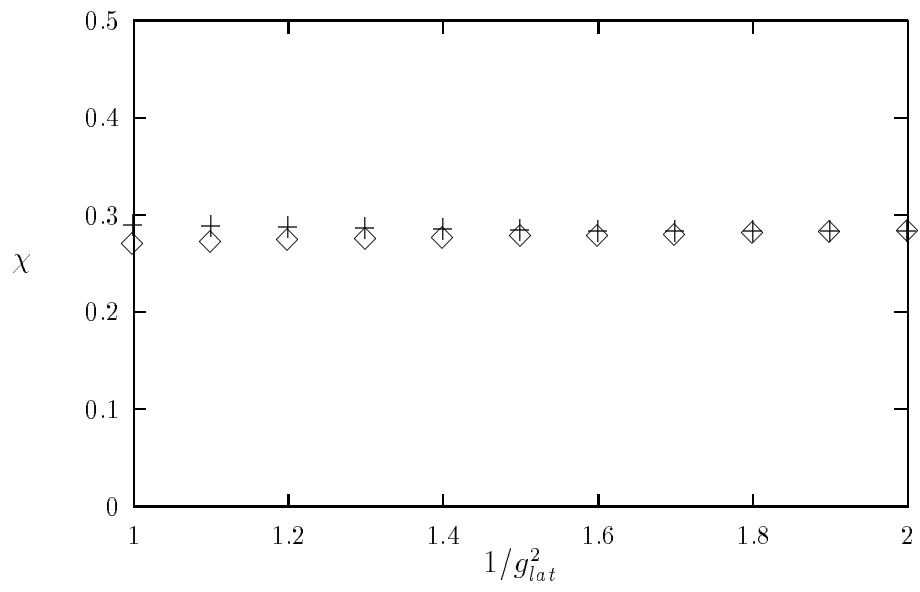
\end{document}